\documentclass[aps,prl,showpacs,citeautoscript,superscriptaddress,longbibliography,notitlepage,twocolumn]{revtex4-1}
\usepackage{braket}
\usepackage{graphicx}  
\usepackage{epstopdf}
\usepackage{dcolumn}   
\usepackage{bm}        
\usepackage{amsmath}
\usepackage{amssymb}   
\usepackage{wrapfig}
\usepackage{array}
\usepackage[caption=false]{subfig}
\usepackage[usenames, dvipsnames]{color}
\usepackage{times, verbatim}
\newcommand{\bld}[1]{\textbf{#1}}

\newcommand{\arddthree}[3]{a^\dagger_{\textbf{r}_{#1}+\textbf{d}_{#2}-\textbf{d}_{#3}}}
\newcommand{\brddtwo}[2]{b^\dagger_{\textbf{r}_{#1}+\textbf{d}_{#2}}}
\newcommand{\ardd}[1]{a^\dagger_{\textbf{r}_{#1}}}

\newcommand{\arone}[1]{a_{\textbf{r}_{#1}}}

\newcommand{\brtwo}[2]{b_{\textbf{r}_{#1}+\textbf{d}_{#2}}}

\begin{document}
\title{Fermionic Chern insulator from twisted light with linear polarization}
\author{Utso Bhattacharya}
\email{utso.bhattacharya@icfo.eu}
\affiliation{ICFO-Institut de Ciencies Fotoniques, The Barcelona Institute of Science and Technology, Av. Carl Friedrich Gauss 3 08860 Castelldefels (Barcelona), Spain}
\affiliation{Max-Planck-Institut für Quantenoptik, D-85748 Garching, Germany}
\author{Swati Chaudhary}
\affiliation{Institute of Quantum Information and Matter and Department of Physics, California Institute of Technology, Pasadena, CA 91125, USA}
\author{Tobias Grass}
\affiliation{ICFO-Institut de Ciencies Fotoniques, The Barcelona Institute of Science and Technology, Av. Carl Friedrich Gauss 3 08860 Castelldefels (Barcelona), Spain}
\author{Allan S. Johnson}
\affiliation{ICFO-Institut de Ciencies Fotoniques, The Barcelona Institute of Science and Technology, Av. Carl Friedrich Gauss 3 08860 Castelldefels (Barcelona), Spain}
\author{Simon Wall}
\affiliation{ICFO-Institut de Ciencies Fotoniques, The Barcelona Institute of Science and Technology, Av. Carl Friedrich Gauss 3 08860 Castelldefels (Barcelona), Spain}
\affiliation{Department of Physics and Astronomy, Aarhus University, Ny Munkegade 120, 8000 Aarhus C, Denmark}
\author{Maciej Lewenstein}
\affiliation{ICFO-Institut de Ciencies Fotoniques, The Barcelona Institute of Science and Technology, Av. Carl Friedrich Gauss 3 08860 Castelldefels (Barcelona), Spain}
\affiliation{ICREA, Pg. Lluis Companys 23, 08010 Barcelona, Spain}

\begin{abstract}
The breaking of time-reversal symmetry is a crucial ingredient to topological bands. It can occur intrisically in materials with magnetic order, or be induced by external fields, such as magnetic fields in quantum Hall systems, or circularly polarized light fields in Floquet Chern insulators. Apart from polarization, photons can carry another degree of freedom, orbital angular momentum, through which time-reversal symmetry can be broken. In this Letter, we pose the question whether this property allows for inducing topological bands via a linearly polarized but twisted light beam. To this end, we study a graphene-like model of electrons on a honeycomb lattice interacting with a twisted light field. To identify topological behavior of the electrons, we calculate their local markers of Chern number, and monitor the presence of in-gap edge states. Our results are shown to be fully analogous to the behavior found in paradigmatic models for static and driven Chern insulators, and realizing the state is  experimentally straightforward. With this, our work establishes a new mechanism for generating Fermionic topological phases of matter that can harness the central phase singularity of an optical vortex beam.
\end{abstract}

\maketitle

In addition to spin angular momentum (SAM), light can also carry orbital angular momentum (OAM) \cite{PhysRevA.45.8185,torres2011twisted}, giving rise to a huge variety of novel opportunities: OAM transfer via light-matter interactions can allow dipole-forbidden atomic transitions, as experimentally demonstrated for a system of trapped ions \cite{schmiegelow16} and theoretically shown for semiconducting and semimetallic materials \cite{quinteiro10,farias13}. It can also be used to generate topological defects in atomic \cite{andersen2006} or excitonic \cite{kwon19} condensates, and OAM beams have already found new applications \cite{shen19} in a variety of fields which include optical communication \cite{wang12,bozinovic13}, quantum information \cite{nicolas14}, cosmology \cite{tamburini11}, and attophysics \cite{kong17,gauthier17}. Despite these advances, the possibility of impinging these beams to create Fermionic topological phases of matter has not yet been explored. 

A paradigmatic model for topological matter \textbf{was} first proposed by F.D.M. Haldane \cite{PhysRevLett.61.2015}: The Haldane model considers electrons on a graphene-like lattice, and by incorporating staggered magnetic fluxes threading through every plaquette of the lattice, the model achieves breaking of the time-reversal symmtery (TRS), while translational invariance remains intact and nearest neighbor hoppings doesn't pick up any magnetic flux. The second nearest neighbor hoppings, however, pick up a net flux which is opposite for the two sublattices.  As can be inferred from the Chern number which becomes $\pm 1$, these complex hoppings of second neighbor electrons create a topological Chern insulator. With this, the Haldane model has provided the conceptual mechanism for theoretical and experimental research exploring various topologically distinct phases of matter. The tremendous importance of this scheme lies in the fact that now (anomalous) quantum Hall effect can appear as an intrinsic property of a band structure, rather than being caused by an external magnetic field. In the past few years, the challenging physical realization of the Haldane model has been achieved in different systems, including magnetically ordered topological insulators \cite{chang13}, and hexagonal arrays of helical waveguides \cite{rechtsman13}. In addition to these examples where time-reversal symmetry is broken by intrinsic properties of the real or synthetic material, there have also been realizations with cold atoms in graphene-like optical lattices \cite{jotzu2014experimental}, or more recently in a real graphene lattice \cite{mciver20}, where time reversal symmetry is broken by driving the system with circularly polarized light.

Such a realization of the Haldane model using Floquet engineering was first proposed by Oka and Aoki \cite{oka2009}. They showed that irradiating a  monolayer of graphene with circularly polarized (CP) light can result in the creation of a Floquet induced Haldane mass. The spin polarization of the light introduces a chirality or a handedness by breaking the TRS of the two band system. The two eigenstates of the Floquet Hamiltonian then acquire non-zero Chern numbers which in turn leads to a non-trivial band topology. The spin polarization or the spin angular momentum (SAM) of CP light being odd under TRS is in fact responsible for the generation of a Haldane mass in the Floquet system. The CP light used there carries SAM, but no net OAM. 

In this Letter we address, and answer affirmatively, the question whether, instead of using CP light with zero net OAM, it is possible to use linearly polarized light (zero SAM), but with some integer OAM, to create a Floquet Chern insulator. Considering the vast applicability of OAM light and the fact that Haldane model serves as a building block for many other topological phases, our work provides a major stepping stone to harness their combined potential through Floquet engineering.

We begin with the tight binding model of graphene. It consists of a monolayer of hexagonal lattice of carbon atoms with two $\pi$-bands \cite{castro_neto09}. Its Hamiltonian $H_0$ is well modeled by nearest neighbor hopping of electrons between two sub-lattices A and B:
\begin{eqnarray}\label{eq_ham}
H_0=&-t_0&\sum_{\bld{r}_i,j={1,2,3}}\left(\brddtwo{i}{j}\arone{i}+\ardd{i}\brtwo{i}{j}\right)\nonumber\\
&+&\Delta_0\sum_{\bld{r}_i}\left(\ardd{i}\arone{i}-\brddtwo{i}{1}\brtwo{i}{1}\right).
\end{eqnarray} 
The Fermionic operators $c_{\bf r} (c_{\bf r}^\dagger)$ annihilates (creates) an electron at position $\mathbf{r}$ with $c=a,b$ for sublattices A and B, respectively. The vectors $\mathbf{r}_i$ span sublattice A, and $\bld{d}_i$ are the position vectors to the three nearest neighbors around each A site of the hexagonal plaquette \footnote{\label{SM} See Supplemental Material which describes the used methods, and which contains reference to Ref. \cite{Eckardt_2015}.}.
The first term in $H_0$ describes the hopping of electrons between the two sublattices A and B with tunneling strength $t_0$, and the second term accounts for a staggered sublattice symmetry breaking mass $\Delta_0$. 

Under periodic boundary conditions, the system is translationally invariant, and for $\Delta_0=0$, it has a gapless spectrum at some isolated points in the quasi-momentum space; the quasi-momentum being a good quantum number due to translational invariance. Such points are called Dirac points (DP) due to their massless relativistic dispersion at very low energies. 
These gapless points are protected by three symmetries: chiral (same mass of the two sublattices), time-reversal, and crystal symmetry $C_3$. However, placing the monolayer on a substrate such as  hexagonal Silicon Carbide or hexagonal Boron Nitride introduces  an energy difference between the sublattices \cite{zhou07}, described by the Semenoff mass term \cite{semenoff84}, which breaks the sublattice or chiral symmetry. Nonetheless, such finite value of $\Delta_0$ only results in a gap that is topologically trivial, due to the presence of TRS. The trivial topology is evidenced by the zero value of a topological invariant called the Chern number calculated over the filled bands of the system. Therefore, to realize a topologically non-trivial insulator, TRS must be broken.

  We now consider two situations where we irradiate a graphene monolayer with: (a) a circularly polarized light with zero OAM and (b) a linearly polarized vortex beam with finite OAM. In this work, we compare these two situations and demonstrate that even the linearly polarized OAM beam can make the system topologically non-trivial.  Here we use the velocity gauge to incorporate the effects of the two periodic drives. The electric field of light $\mathbf{E}(t)$ in the Coulomb gauge is related to the vector potential via the relation $\mathbf{E}(t)=-\frac{\partial \mathbf{A}(t)}{\partial t}$; we also set $\hbar = 1$ and speed of light $c = 1$. We consider the effect of this vector potential on the hopping term by making Peierls substitution \cite{Note1}.
To analyze the two cases mentioned above, we employ Floquet theory and use van Vleck high frequency expansion \cite{bukov15} to derive an effective Floquet Hamiltonian. Numerically, using exact diagonalization, as well as a semi-analytic construction, we study the Floquet Hamiltonian in both situations. 

We first discuss the case of graphene irradiated with CP light which is characterized by a vector potential:
\begin{equation}\label{eq_ac}
\textbf{A}=A_0\cos\omega t~\hat{x}\pm A_0\sin\omega t~\hat{y}
\end{equation}
where $\hat{x}, \hat{y}$ are orthonormal unit vectors in Cartesian coordinates, $\pm$ stands for right- or left -circular polarisation of the light field, $\omega$ is the frequency, and $A_0$ is the amplitude of the vector potential which in the dipole approximation, is considered spatially uniform over the whole lattice. It can be shown that this drive generates complex next nearest neighbor (NNN) hopping terms with opposite sign for $A$ and $B$ sites. As a result of this term, the system becomes  equivalent to the static Haldane Chern insulator. In this case, the NNN hopping gives rise to a TRS breaking mass term $m_\text{CP}\propto\frac{t_0^2(J_1(A_ 0))^2}{\omega}$ for each valley in the reciprocal space,  where $J_1$ denotes the first Bessel function. This mass term competes with the Semenoff mass term to push the system to a topological phase with non-zero Chern invariant. For large frequencies and small electric field amplitudes $E_\text{amp}  = i \omega A_0$, this terms goes as $\frac{E_\text{amp}^2}{\omega^3}$, see \cite{Note1}.

We now turn our attention to the second scenario, where we shine linearly polarized OAM beam on graphene.
For OAM drive, we use a vector potential given by:
\begin{equation}\label{eq_ao}
\textbf{A}(x,y)=\left(A_x(r)e^{il\phi}e^{i\omega t}+A_x^*(r)e^{-il\phi}e^{-i\omega t}\right)\hat{x}
\end{equation}
where $\phi=\text{tan}^{-1}(\frac{y-y_0}{x-x_0})$ and in the present case, we chose its center at $(x_0,y_0)=(0,0)$ so that it doesn't lie directly on any lattice site. We immediately note that the vector potential is position dependent and thus breaks translational invariance of the lattice. As a result of this spatial dependence,  quasi-momenta $k_x, k_y$ are no longer good quantum numbers. Therefore, to study the system a real space analysis is imperative.
The amplitude $A_x(r)$ is a function of radial position $r$  with a singularity at $r=0$, as for instance in a Laguerre-Gaussian or a Bessel beam, 
but for now, we choose a constant amplitude $A_0$ for $r\ne0$ so that we can solely focus on the phase of the light which generates the OAM. Again, we include the vector potential via Peierls substitution  to derive a Floquet Hamiltonian \cite{Note1}. However, in contrast to the CP drive, in this case, both the NNN hopping amplitude and phase in the effective Hamiltonian are position dependent.

 \begin{figure*}
 	\includegraphics[scale=0.29]{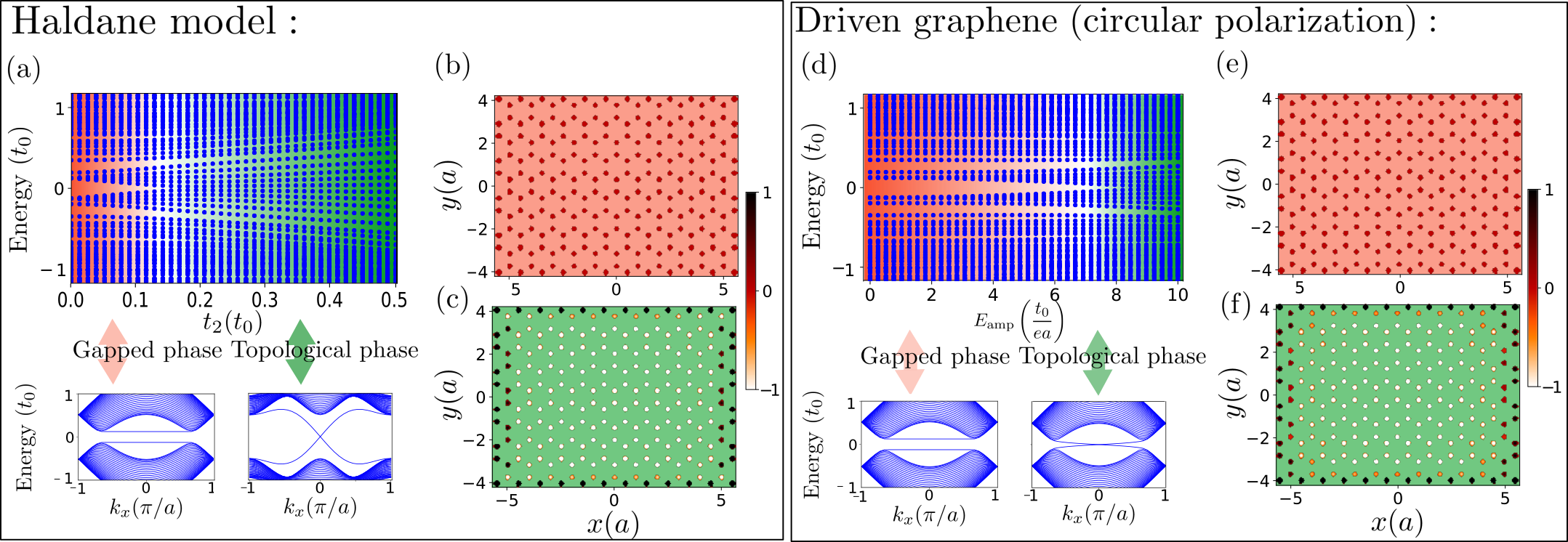}
	\caption{ {\bf Static vs. driven Chern insulator.} 
	(a) Spectra for Haldane model ($\Delta_0=0.125 t_0$):  Energy vs. next-nearest neighbor hopping $t_2$ in a system with open boundary (upper plot), and energy vs. wavevector $k_x$ in a system with periodic boundary (lower plots), for gapped and topological regime respectively. The topological regime is characterized by edge states crossing the band gap. (b) Local Markers of Chern number (LMCN) in the trivially gapped regime (graphene Hamiltonian), and (c) LMCN in the non-trivial regime (Haldane Hamiltonian). 
	(d) Floquet spectra for graphene driven by circularly polarized light ($\Delta_0=0.125 t_0$, $\hbar\omega=15 t_0$): Energy vs. driving amplitude $E_{\rm amp}$ in a system with open boundary (upper plot), and energy vs. wavevector $k_x$ in a system with periodic boundary (lower plots), for gapped and topological regime respectively. (e) LMCN in the trivial regime, (f) LMCN in non-trivial regime, of driven graphene.
	}
	\label{Haldanenew}
\end{figure*}

Generally, a direct measure of the bulk topology of a system is the Chern invariant, which is usually defined as an integral over the Berry curvature of a band, and thus requires the presence of translational invariance. Since the present scenario lacks translational invariance, we cannot calculate the quasimomentum-space bulk Chern number, and hence we rely on the real-space formulation discussed in Ref. \cite{PhysRevB.84.241106,PhysRevB.100.214109}. In this formalism, one can infer the Chern number of the system from the macroscopic average of a gauge invariant microscopic function known as local marker of Chern number (see \cite{Note1}). We quickly present the result of the topological invariant from the local markers of Chern number (LMCN) for the following three situations: (i) trivially gapped graphene, (ii) Haldane model, (iii) CP driven graphene  (non-zero SAM, zero OAM), and then compare the results with that of (iv) OAM driven graphene (zero SAM, non-zero OAM). For all cases, we consider a lattice under open boundary condition (OBC) in both directions, and we plot the LMCN for every point in the lattice: 
Fig. \ref{Haldanenew}(b) shows the result for case (i), with the local Chern number being zero everywhere in the bulk and at the edges. This correctly captures the fact that TRS is intact, and the energy gap due to broken sublattice symmetry leaves the system topologically trivial.
In contrast, Figs.~\ref{Haldanenew}(c) and (f), which correspond to the cases (ii) and (iii) of broken TRS, clearly demonstrate topologically non-trivial behavior. Specifically, the LMCN exhibits edge states (in black as shown in the colorbars alongside the figure) with averaged bulk Chern number $+1$.
We also note that the edge LMCN and the averaged bulk Chern number have equal and opposite contributions as expected according to Ref. \cite{PhysRevB.84.241106}. Furthermore, on reversing the sign of the mass term in the static case or the polarisation of light in the driven scenario, the edge (and bulk) value of LMCN reverses as expected (see \cite{Note1}). These results establish that LMCN are an excellent means of identifying topological phases in the absence of translational invariance.

\begin{figure*}
	\centering
	\includegraphics[scale=0.27]{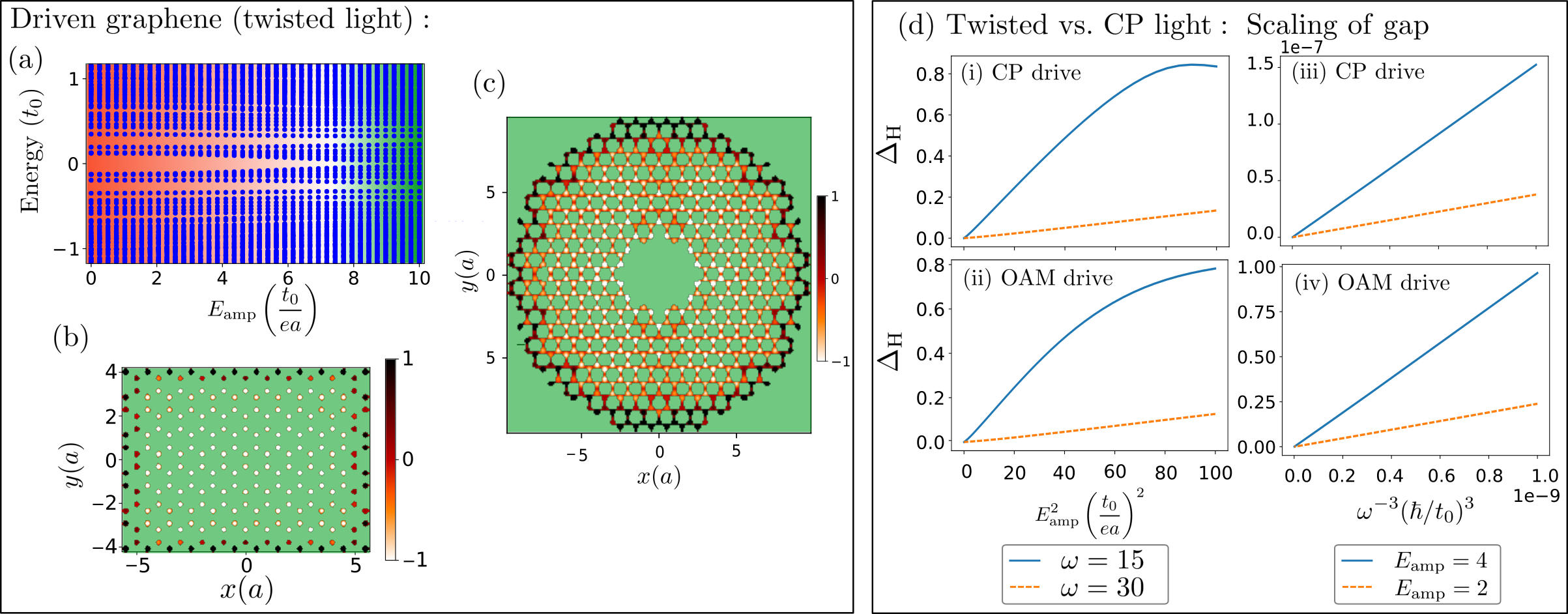}
	\caption{
		{\bf Chern insulator due to OAM drive.} (a) Floquet spectrum: Energy vs. driving amplitude $E_{\rm amp}$ in a system with open boundary ($\Delta_0=0.125 t_0$, $\hbar\omega=15 t_0$): Closing of the band gap due to strong driving indicates the transition into a non-trivial phase. (b,c) Local Chern numbers: Topologically non-trivial behavior of OAM driven graphene is evident from the non-zero local Chern numbers, in a rectangular sample (b), and a corbino-shaped geometry (c). (d) Scaling of the Haldane like gap, $\Delta_H$ with driving amplitude and  frequency for both CP-driven and OAM-driven graphene. Here, $\Delta_H= \frac{2\Delta_0-E_\text{gap}}{2\Delta_0}$, where $E_\text{gap}$ is the band gap shown in (a).}
	\label{OAM drivenew}
\end{figure*}

 We are now set to apply this tool to case (iv), our prime case of interest.  Fig. \ref{OAM drivenew}(b) shows the corresponding LMCN plot for a rectangular system, while Fig. \ref{OAM drivenew}(c) shows the LMCN for a sample with Corbino geometry.
In the rectangular system, we have assumed that the light beam is of constant intensity. As this assumption does not hold in the vicinity of the optical vortex, we have also considered a sample in a Corbino geometry, in which the optical vortex of a Bessel beam is placed at the center of the annular hole albeit making sure that the intensity from the beam reaches its peak value between the two edges of the sample. 
Both plots, Figs. \ref{OAM drivenew}(b) and (c), conclusively reveal that the system hosts chiral topological edge states. In case of the Corbino sample, 
the inner circular edge of the sample has a non-zero integer quantized LMCN with value $-1$ (white) whereas the outer edge of the sample has a LMCN with value $+1$ (black).
 Similarly, in case of the rectangular system, the LMCN at the sample edge averages to $+1$ (black). Furthermore, in both cases, the local Chern number changes sign along the two edges on reversing the OAM of the beam, indicating that the chirality of these states is due to the OAM of the beam.
This conclusion is also corroborated by the constant intensity case (rectangular sample), which reveals that the phase winding of the beam is sufficient to generate the Chern insulating behavior.
We now want to take a deeper look at the topological edge states in the three cases (ii)--(iv) discussed above. To this end, under complete OBC, we plot the energy spectra against the parameter which controls the complex NNN hopping:  In the static Haldane model, this is simply $t_2$, whereas for the driven cases it is the amplitude $E_\text{amp}$ of the light field. All spectra, shown in Fig.~\ref{Haldanenew}(a),~\ref{Haldanenew}(d), and ~\ref{OAM drivenew}(a), have one feature in common: 
 The energy gap due to a finite Semenoff mass closes when the control parameter becomes sufficiently strong, which distinguishes a gapped phase (shaded in red) from a gapless (at edge) phase (shaded in green).
The observed gap closing is due to the occurence of an in-gap state, which can easily be visualized for case (ii) and case (iii). In these cases, the presence of translational symmetry allows us to apply PBC along the $x$-direction, and the lower plots in Figs.~\ref{Haldanenew}(a) and (d) show the resulting energy dispersion in both the gapped and the gapless regime.
We observe that the vanishing of the gap is due to a single energy level which connects one valley of the  conduction band to the opposite in the valence band. 
In contrast, in the gapped regime, the energy dispersion exhibits topologically trivial edge states which connect the Dirac modes within each band, but without crossing the gap between conduction and valence band. Due to the lack of translational invariance an analogous plot for the quasienergy spectra against a conserved quasi-momentum is not possible in the OAM driven scenario, but the appearance of in-gap states is seen from Fig.~\ref{OAM drivenew}(a), and hints for a non-trivial topological phase with chiral edge states.

We finally want to reveal the order of the amplitude and frequency of OAM light required to produce such a Floquet Chern insulator. A semi-analytical approach is adopted, in which we construct an effective Floquet Hamiltonian once again via a van Vleck transformation, but this time in the limit where only the leading order term is retained (see \cite{Note1}), that is, the term which scales with $E_\text{amp}^2/\omega^3$, or the single photon sector. We use numerical diagonalization to confirm that the quasienergy spectrum around zero energy for this Hamiltonian and with that obtained from exact numerical approach (see Fig. 2 in \cite{Note1}) match at high frequencies and small amplitudes of the light field. Most importantly, the Floquet Chern insulator from OAM possesses a mass gap, $\Delta_H$ which scales as $E_\text{amp}^2/\omega^3$,  and thus it exhibits exactly the same quantitative behavior as the CP light driven system (as shown in Fig.\ref{OAM drivenew}(d)). The identical topological behavior of the eigenstates captured through the LMCN and the correspondence of the zero energy spectrum is suggestive of a topological equivalence between the  effective Hamiltonian and the Hamiltonian of static Haldane model, which are formally very similar, albeit with position-dependent phases along the NNN electron hoppings in the effective Hamiltonian. 

Experimentally generating such OAM beams with high field strengths required to observe the Chern insulator behavior is now routine, both in the laboratory (UV-THz) and at facilities (XUV-X-ray). Realizing the geometry simulated here, with uniform intensity outside of the central singularity located in a central aperture, is less standard but not significantly more challenging. The intensity full-width half maximum of the central minimum in OAM beams can be as low as 0.3 wavelengths \citep{Torok2004}, and with appropriate phase masking the intensity can be made approximately uniform outside of this region without affecting the carried OAM. For high intensity X-rays at X-ray free electron lasers this is already sufficient to address graphene in the geometry shown in Fig.2(c), which is limited to a radial size of 10 lattice constants. Longer wavelengths require larger samples, or alternatively, larger lattice constants. Artificial forms of graphene can have lattice constants up to 100 nm \cite{polini2013artificial}. In all cases the topologically non-trivial state can be detected via the emergence of an anomalous Hall current as in Ref. \onlinecite{mciver20}. Since the gap is found to saturate with high field strengths (Fig.\ref{OAM drivenew}(d)) it should also be possible to decrease the effective width of the minimum by going to higher field strengths, analogously to STED microscopy \citep{Klar2001}, allowing optical fields to produce the effect directly in graphene.

Circularly polarized light pulses can induce the Floquet Chern phase in monolayer graphene \cite{mciver20}. In this Letter, we have demonstrated that, amazingly, the Floquet Chern phase  can also be induced via linearly polarized light with non-zero OAM. To this aim, we calculated the Local Markers of Chern Number and monitored the presence of in-gap edge states. This opens new paths for the light induced  generation of topological states in solid state systems consisting of monolayers, bilayers, etc. It would be interesting to carry over these ideas to quantum  simulator platforms, such as cold atoms. Another interesting possibility could be to  combine OAM and CP to create and impinge on the substrate even more ``exotic'' topological light pulses, such as the ones carrying topological knots \cite{pisanty2019knotting,PhysRevLett.122.203201} , or bi-chromatic fields which can be used, for instance, to generate XUV harmonics with a self-torque, i.e. time-dependent OAM \cite{rego2019generation}.

\begin{acknowledgments} 
 We thank Gil Refael for enlightening discussions. U.B.,T.G., and M.L. acknowledge the ERC AdG NOQIA, State Research Agency AEI (“Severo Ochoa” Center of Excellence CEX2019-000910-S, Plan National FIDEUA PID2019-106901GB-I00/10.13039 / 501100011033, FPI, QUANTERA MAQS PCI2019-111828-2 / 10.13039/501100011033), Spanish Ministry MINECO (National Plan 15 Grant: FISICATEAMO No.FIS2016-79508-P, SEVERO OCHOA No. SEV-2015-0522,FPI), European Social Fund, Fundaci Cellex, Fundaci Mir-Puig, Generalitat de Catalunya (AGAUR Grant No. 2017SGR 1341, CERCA program, QuantumCAT \ U16-011424, co-funded by ERDF Operational Program of Catalonia 2014-2020), EU Horizon 2020 FET-OPEN OPTOLogic (Grant No 899794), and the National Science Centre, Poland-Symfonia Grant No. 2016/20/W/ST4/00314, Marie Sk\l odowska-Curie grant STREDCH No 101029393, “La Caixa” Junior Leaders fellowships (ID100010434),  and EU Horizon 2020 under Marie Sk\l odowska-Curie grant agreement No 847648 (LCF/BQ/PI19/11690013, LCF/BQ/PI20/11760031,  LCF/BQ/PR20/11770012). S.C. acknowledges support from the Institute of Quantum Information and Matter, an NSF Frontier center funded by the Gordon and Betty Moore Foundation. T.G. acknowledges financial support from a fellowship granted by la Caixa Foundation (ID100010434, fellowship code LCF/BQ/PI19/11690013). U.B. acknowledges Cellex-ICFO-MPQ fellowship funding. A.S.J. acknowledges funding from Marie Skodowska-Curie grant agreement No. 754510 (PROBIST). All numerical simulations were performed using Kwant\cite{Groth_2014}. 
\end{acknowledgments}

\section{Supplemental Material: Fermionic Chern insulator from twisted light with linear polarization}

We describe the method to obtain effective Floquet Hamiltonians in the driven system, and provide the definition of the local markers of Chern number used to topologically characterize the phases.

\subsection{General form of Floquet Hamiltonian}
The general form of the Floquet Hamiltonian which has been used through out this work is given by
\begin{equation}
\begin{split}
H=&\sum_{n=-N}^{n=N}(H_0\otimes\ket{n}\bra{n}+n\omega\mathbb{I}\otimes\ket{n}\bra{n}\\&+H_m\otimes\ket{n}\bra{n+m}+H_m^\dagger\otimes\ket{n}\bra{n-m})
\end{split}
\end{equation}
where $H_0$ is the undriven Hamiltonian for graphene, and $\ket{n}$ denotes  denotes the photon degree of freedom~\cite{Eckardt_2015}. The elements $H_m$ are Fourier coefficients of the Hamiltonian $H_m = \int_0^T {\rm d}t H(t) e^{-im\omega t}$.

 In matrix notation, the Floquet Hamiltonian reads

  \begin{equation}
 \begin{pmatrix}
 H_0+N\omega &H_1&\cdots& & H_{2N} \\
 H_1^\dagger&\ddots&& & \\
 \vdots  &  & H_0  & & \vdots\\ 
 & & &  \ddots & H_1 \\
 H_{2N}^\dagger & &  \cdots & H_1^\dagger& H_0 - N\omega \\
 \end{pmatrix}
 \label{FT}
 \end{equation}
 
By numerical diagonalization we have obtained the spectrum and eigenstates of the system by truncating the Floquet Hamiltonian beyond $N=1$ sector. We have verified that increasing $N$ does not result in any significant change in the LMCN or the spectra.

\begin{figure}[h]
	\includegraphics[scale=0.5]{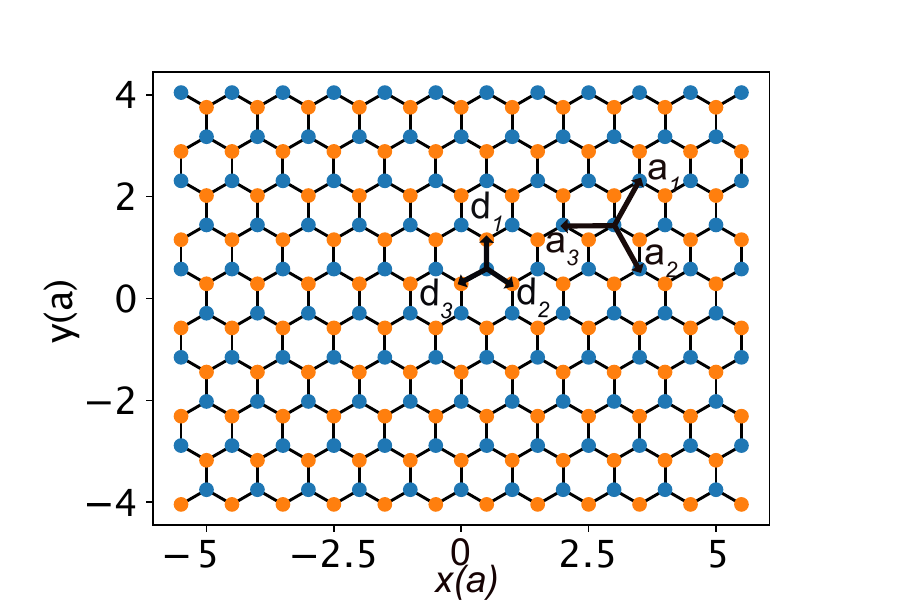}
	
	\caption{Graphene lattice.}
	\label{system}
\end{figure}

The graphene Hamiltonian $H_0$ consists of nearest-neighbor hopping on a lattice as shown in Fig.\ref{system}, and, potentially, a substrate-induced mass term between the two sublattices. The vectors connecting one site to its three nearest neighbors are given by:
\begin{equation}
\begin{split}
\textbf{d}_1=&a(0,1),\,\textbf{d}_2=a\left(\cos\left(\frac{\pi}{6}\right),-\sin\left(\frac{\pi}{6}\right)\right),\,\\& \textbf{d}_3=a\left(-\cos\left(\frac{\pi}{6}\right),-\sin\left(\frac{\pi}{6}\right)\right),
\end{split}
\end{equation}
where $a$ is the lattice constant of the underlying triangular lattice.

The periodic driving of the graphene lattice, either with circularly polarized Gaussian light or with linearly polarized OAM light, is incorporated via the Peierls' substitution:
\begin{equation}
t_{\textbf{r}_i\textbf{r}_j}\rightarrow e^{i\int_{\textbf{r}_i}^{\textbf{r}_j}\textbf{A}\cdot d\textbf{r}}t_{\textbf{r}_i\textbf{r}_j}\rightarrow e^{i\left(\int_0^{\textbf{r}_i}\textbf{A}\cdot d\textbf{r}-\int_0^{\textbf{r}_j}\textbf{A}\cdot d\textbf{r}\right)}t_{\textbf{r}_i\textbf{r}_j}
\end{equation}
where $\textbf{A}$ is the vector potential of the light used to irradiate the sample.

\subsection{Effective Floquet Hamiltonian for CP drive}
For the CP drive, the vector potential reads $\textbf{A}=A_0\cos\omega t\hat{x}+A_0\sin\omega t\hat{y}$ and thus Peierls' phase takes the following form
\begin{equation}
e^{i\left(\int_0^{\textbf{r}_i}\textbf{A}\cdot d\textbf{r}-\int_0^{\textbf{r}_j}\textbf{A}\cdot d\textbf{r}\right)}=e^{i\alpha\sin(\omega t+\beta)}
\end{equation}
where $\beta$ depends on $\textbf{r}_i$ and $\textbf{r}_j$. Now, we express the above expression using Jacobi-Anger expression:
\begin{equation}
e^{i\alpha\sin(\omega t+\beta)}=\sum_n J_n(\alpha)e^{i n\beta}e^{in\omega t}.
\end{equation}

This allows us to find the Fourier components $H_n$ which occur in Eq.(\ref{FT}). Specifically, the real space $H_{\pm 1}$ are:
\begin{equation}
H_1=t_0\sum_{\bld{r}_i,j={1,2,3}}J_1(A)e^{i\phi_j}\left(\brddtwo{i}{j}\arone{i}-\ardd{i}\brtwo{i}{j}\right)
\end{equation}
and
\begin{equation}
H_{-1}=t_0\sum_{\bld{r}_i,j={1,2,3}}J_{-1}(A)e^{-i\phi_j}\left(\brddtwo{i}{j}\arone{i}-\ardd{i}\brtwo{i}{j}\right)
\end{equation}
where
$\phi_1=\pi,\,\phi_2=\frac{\pi}{3},\,\phi_3=-\frac{\pi}{3}$, and $A=\frac{eEa}{\omega}$. The $\phi$ phase is obtained via a Peierl's substitution of the vector potential. It can be shown that the commutation 
\begin{equation}
\begin{split}
[H_1,H_{-1}]=t_0^2(J_1(A))^2\sum_{\textbf{r}_i,\bld{d}_k-\bld{d}_j={\textbf{a}_1,\textbf{a}_2,\textbf{a}_3}}2i\sin(\phi_k-\phi_j)\\ \left(\brddtwo{i}{k}\brtwo{i}{j}-\arddthree{i}{k}{j}\arone{i}\right)+\text{h.c}
\end{split}
\end{equation}
where $\textbf{a}_3=-\textbf{a}_1-\textbf{a}_2$ and $\textbf{a}_1,\textbf{a}_2$ are two lattice vectors for the given honeycomb lattice (see Fig.3). The hopping becomes complex, but remains independent from position, as $\phi_1-\phi_2=\phi_2-\phi_3=\phi_3-\phi_1=\frac{2\pi}{3}$. Thus, to lowest order (i.e. in the high-frequency regime), the effective Floquet Hamiltonian becomes identical to the static Haldane model with $t_2\propto t_0^2\frac{E}{w^3}$.

\subsection{Effective Floquet Hamiltonian for OAM drive}
In the case of OAM drive, the vector potential $
\textbf{A}(x,y)=\left(A_x(r)e^{il\phi}e^{i\omega t}+A_x^*(r)e^{-il\phi}e^{-i\omega t}\right)\hat{x}
$, and after Peierls' phase substitution, we get $H_{\pm 1}$:
\begin{equation}
\begin{split}
H_1=t_0\sum_{\bld{r}_i,j={1,2,3}}J_1(A_{ji})e^{i\phi_{ij}}\left(\brddtwo{i}{j}\arone{i}-\ardd{i}\brtwo{i}{j}\right)
\end{split}
\end{equation}
with
\begin{equation}
H_{-1}=t_0\sum_{\bld{r}_i,j={1,2,3}}J_{-1}(A_{ij})e^{-i\phi_{ij}}\left(\brddtwo{i}{j}\arone{i}-\ardd{i}\brtwo{i}{j}\right)
\end{equation}
where 
\begin{equation}
A_{ij}=\frac{e}{\omega}\left|\int_{0}^{\textbf{r}_i+\textbf{d}_j}e^{il\theta}\textbf{E}\cdot d\textbf{r}-\int_{0}^{\textbf{r}_i}e^{il\theta}\textbf{E}\cdot d\textbf{r}\right|
\end{equation}
and 
\begin{equation}
\phi_{ij}=\text{Arg}\left(\int_{0}^{\textbf{r}_i+\textbf{d}_j}e^{il\theta}\textbf{E}\cdot d\textbf{r}-\int_{0}^{\textbf{r}_i}e^{il\theta}\textbf{E}\cdot d\textbf{r}\right).
\end{equation}
The first contribution $H^{(1)}$ comes out to be
\begin{equation}
\begin{split}
H^{(1)}&=[H_1,H_{-1}]=t_0^2\sum_{\textbf{r}_i,\bld{d}_k-\bld{d}_j={\textbf{a}_1,\textbf{a}_2,\textbf{a}_3}}iJ_1(A_{ik})J_1(A_{ij})\\& \left(\sin\left(\phi_{ik}-\phi_{ij}\right)\brddtwo{i}{k}\brtwo{i}{j}\right)-t_0^2\sum_{\textbf{r}_i,\bld{d}_k-\bld{d}_j={\textbf{a}_1,\textbf{a}_2,\textbf{a}_3}}\\&iJ_1(A_{ik})J_1(A_{ij})\sin(\phi_{ik}-\phi_{(i+\alpha) j})\arddthree{i}{k}{j}\arone{i}+\text{h.c}
\end{split}
\label{commutation_OAM}
\end{equation}
where $\textbf{r}_{i+\alpha}=\textbf{r}_i+\textbf{a}_\alpha$. Again, in the high-frequency regime the effective OAM Hamiltonian is given by
\begin{equation}
H_\text{eff}=H_0+\frac{H^{(1)}}{\omega}+O\left(\frac{1}{\omega^2}\right).
\end{equation}
As for CP driven case discussed above, $H^{(1)}$ introduces complex hopping between next-nearest neighbor sites, so this analysis reveals the formal ressemblance of the two cases.
However, in the OAM driven case, the phase and amplitude of the next-nearest neighbor hopping depends on position in a complicated manner. This impedes an analytical solution, in particular as periodic boundary conditions cannot be applied. For this reason, information about the topology of the system has been obtained from the local marker of the Chern number \cite{PhysRevB.84.241106,PhysRevB.100.214109} which we explicitly define in next section.

\begin{figure}[t]
	\includegraphics[scale=0.5]{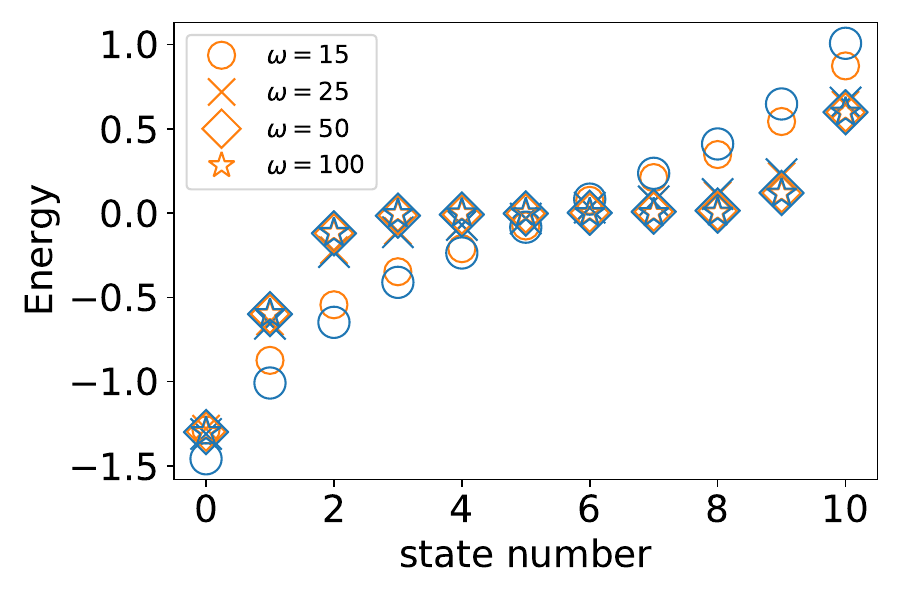}
	\caption{Spectrum around zero energy as calculated from the full Floquet Hamiltonian $H_F$ (orange) and the effective Floquet Hamiltonian from leading order in high-frequency expansion $H_\text{new}$ (blue). }
	\label{comparison}
\end{figure}
In order to evaluate the validity of the high-frequency expansion we compare the spectrum of the full Floquet Hamiltonian $H_F$ from Eq.~(\ref{FT}) (truncated at $N=1$), and $H_\text{new}$ in Fig.~\ref{comparison}. This confirms the validity of $H_\text{new}$ for driving frequency $\omega \gtrsim 25 t_0$.

%

 
%

\subsection{Local Chern number}
\label{lcn}
To identify topological properties of the system we have used local Chern numbers according to the recipe discussed in Refs.~\cite{PhysRevB.84.241106,PhysRevB.100.214109}. The local Chern numbers are defined as
\begin{equation}
C(\textbf{r}_i)=\frac{2\pi}{A_c}\bra{\textbf{r}_i}[PxP,PyP]\ket{\textbf{r}_i}
\end{equation}
where $P$ is the projection on states with $E>E_F$ given by:
\begin{equation}
P=\sum_{E_\lambda>E_F}\ket{\psi_\lambda}\bra{\psi_\lambda}=\sum_{E_\lambda>E_F}\sum_{\textbf{r}_i,\textbf{r}_j}\ket{\textbf{r}_i}\left<\textbf{r}_i|\psi_\lambda\right>\left<\psi_\lambda|{\textbf{r}_j}\right>\bra{\textbf{r}_j}
\end{equation}
where
$x=\sum_{\textbf{r}_i}x_i\ket{\textbf{r}_i}\bra{\textbf{r}_i}$.

\bibliography{myrefs.bib}

\begin{thebibliography}{35}%
\makeatletter
\providecommand \@ifxundefined [1]{%
 \@ifx{#1\undefined}
}%
\providecommand \@ifnum [1]{%
 \ifnum #1\expandafter \@firstoftwo
 \else \expandafter \@secondoftwo
 \fi
}%
\providecommand \@ifx [1]{%
 \ifx #1\expandafter \@firstoftwo
 \else \expandafter \@secondoftwo
 \fi
}%
\providecommand \natexlab [1]{#1}%
\providecommand \enquote  [1]{``#1''}%
\providecommand \bibnamefont  [1]{#1}%
\providecommand \bibfnamefont [1]{#1}%
\providecommand \citenamefont [1]{#1}%
\providecommand \href@noop [0]{\@secondoftwo}%
\providecommand \href [0]{\begingroup \@sanitize@url \@href}%
\providecommand \@href[1]{\@@startlink{#1}\@@href}%
\providecommand \@@href[1]{\endgroup#1\@@endlink}%
\providecommand \@sanitize@url [0]{\catcode `\\12\catcode `\$12\catcode
  `\&12\catcode `\#12\catcode `\^12\catcode `\_12\catcode `\%12\relax}%
\providecommand \@@startlink[1]{}%
\providecommand \@@endlink[0]{}%
\providecommand \url  [0]{\begingroup\@sanitize@url \@url }%
\providecommand \@url [1]{\endgroup\@href {#1}{\urlprefix }}%
\providecommand \urlprefix  [0]{URL }%
\providecommand \Eprint [0]{\href }%
\providecommand \doibase [0]{http://dx.doi.org/}%
\providecommand \selectlanguage [0]{\@gobble}%
\providecommand \bibinfo  [0]{\@secondoftwo}%
\providecommand \bibfield  [0]{\@secondoftwo}%
\providecommand \translation [1]{[#1]}%
\providecommand \BibitemOpen [0]{}%
\providecommand \bibitemStop [0]{}%
\providecommand \bibitemNoStop [0]{.\EOS\space}%
\providecommand \EOS [0]{\spacefactor3000\relax}%
\providecommand \BibitemShut  [1]{\csname bibitem#1\endcsname}%
\let\auto@bib@innerbib\@empty
\bibitem [{\citenamefont {Allen}\ \emph {et~al.}(1992)\citenamefont {Allen},
  \citenamefont {Beijersbergen}, \citenamefont {Spreeuw},\ and\ \citenamefont
  {Woerdman}}]{PhysRevA.45.8185}%
  \BibitemOpen
  \bibfield  {author} {\bibinfo {author} {\bibfnamefont {L.}~\bibnamefont
  {Allen}}, \bibinfo {author} {\bibfnamefont {M.~W.}\ \bibnamefont
  {Beijersbergen}}, \bibinfo {author} {\bibfnamefont {R.~J.~C.}\ \bibnamefont
  {Spreeuw}}, \ and\ \bibinfo {author} {\bibfnamefont {J.~P.}\ \bibnamefont
  {Woerdman}},\ }\bibfield  {title} {\enquote {\bibinfo {title} {{Orbital
  angular momentum of light and the transformation of Laguerre-Gaussian laser
  modes}},}\ }\href {\doibase 10.1103/PhysRevA.45.8185} {\bibfield  {journal}
  {\bibinfo  {journal} {Phys. Rev. A}\ }\textbf {\bibinfo {volume} {45}},\
  \bibinfo {pages} {8185--8189} (\bibinfo {year} {1992})}\BibitemShut {NoStop}%
\bibitem [{\citenamefont {Torres}\ and\ \citenamefont
  {Torner}(2011)}]{torres2011twisted}%
  \BibitemOpen
  \bibfield  {author} {\bibinfo {author} {\bibfnamefont {J.P.}\ \bibnamefont
  {Torres}}\ and\ \bibinfo {author} {\bibfnamefont {L.}~\bibnamefont
  {Torner}},\ }\href {https://books.google.com/books?id=WK-zHDuHET4C} {\emph
  {\bibinfo {title} {{Twisted Photons: Applications of Light with Orbital
  Angular Momentum}}}}\ (\bibinfo  {publisher} {Wiley},\ \bibinfo {year}
  {2011})\BibitemShut {NoStop}%
\bibitem [{\citenamefont {Schmiegelow}\ \emph {et~al.}(2016)\citenamefont
  {Schmiegelow}, \citenamefont {Schulz}, \citenamefont {Kaufmann},
  \citenamefont {Ruster}, \citenamefont {Poschinger},\ and\ \citenamefont
  {Schmidt-Kaler}}]{schmiegelow16}%
  \BibitemOpen
  \bibfield  {author} {\bibinfo {author} {\bibfnamefont {Christian~T.}\
  \bibnamefont {Schmiegelow}}, \bibinfo {author} {\bibfnamefont {Jonas}\
  \bibnamefont {Schulz}}, \bibinfo {author} {\bibfnamefont {Henning}\
  \bibnamefont {Kaufmann}}, \bibinfo {author} {\bibfnamefont {Thomas}\
  \bibnamefont {Ruster}}, \bibinfo {author} {\bibfnamefont {Ulrich~G.}\
  \bibnamefont {Poschinger}}, \ and\ \bibinfo {author} {\bibfnamefont
  {Ferdinand}\ \bibnamefont {Schmidt-Kaler}},\ }\bibfield  {title} {\enquote
  {\bibinfo {title} {{Transfer of optical orbital angular momentum to a bound
  electron}},}\ }\href {\doibase 10.1038/ncomms12998} {\bibfield  {journal}
  {\bibinfo  {journal} {Nature Communications}\ }\textbf {\bibinfo {volume}
  {7}},\ \bibinfo {pages} {12998} (\bibinfo {year} {2016})}\BibitemShut
  {NoStop}%
\bibitem [{\citenamefont {Quinteiro}\ and\ \citenamefont
  {Tamborenea}(2010)}]{quinteiro10}%
  \BibitemOpen
  \bibfield  {author} {\bibinfo {author} {\bibfnamefont {G.~F.}\ \bibnamefont
  {Quinteiro}}\ and\ \bibinfo {author} {\bibfnamefont {P.~I.}\ \bibnamefont
  {Tamborenea}},\ }\bibfield  {title} {\enquote {\bibinfo {title}
  {{Twisted-light-induced optical transitions in semiconductors: Free-carrier
  quantum kinetics}},}\ }\href {\doibase 10.1103/PhysRevB.82.125207} {\bibfield
   {journal} {\bibinfo  {journal} {Phys. Rev. B}\ }\textbf {\bibinfo {volume}
  {82}},\ \bibinfo {pages} {125207} (\bibinfo {year} {2010})}\BibitemShut
  {NoStop}%
\bibitem [{\citenamefont {Far{\'i}as}\ \emph {et~al.}(2013)\citenamefont
  {Far{\'i}as}, \citenamefont {Quinteiro},\ and\ \citenamefont
  {Tamborenea}}]{farias13}%
  \BibitemOpen
  \bibfield  {author} {\bibinfo {author} {\bibfnamefont {M.~B.}\ \bibnamefont
  {Far{\'i}as}}, \bibinfo {author} {\bibfnamefont {G.~F.}\ \bibnamefont
  {Quinteiro}}, \ and\ \bibinfo {author} {\bibfnamefont {P.~I.}\ \bibnamefont
  {Tamborenea}},\ }\bibfield  {title} {\enquote {\bibinfo {title}
  {{Photoexcitation of graphene with twisted light}},}\ }\href {\doibase
  10.1140/epjb/e2013-40621-2} {\bibfield  {journal} {\bibinfo  {journal} {The
  European Physical Journal B}\ ,\ \bibinfo {pages} {432}} (\bibinfo {year}
  {2013})}\BibitemShut {NoStop}%
\bibitem [{\citenamefont {Andersen}\ \emph {et~al.}(2006)\citenamefont
  {Andersen}, \citenamefont {Ryu}, \citenamefont {Clad{\'e}}, \citenamefont
  {Natarajan}, \citenamefont {Vaziri}, \citenamefont {Helmerson},\ and\
  \citenamefont {Phillips}}]{andersen2006}%
  \BibitemOpen
  \bibfield  {author} {\bibinfo {author} {\bibfnamefont {M.~F.}\ \bibnamefont
  {Andersen}}, \bibinfo {author} {\bibfnamefont {C.}~\bibnamefont {Ryu}},
  \bibinfo {author} {\bibfnamefont {Pierre}\ \bibnamefont {Clad{\'e}}},
  \bibinfo {author} {\bibfnamefont {Vasant}\ \bibnamefont {Natarajan}},
  \bibinfo {author} {\bibfnamefont {A.}~\bibnamefont {Vaziri}}, \bibinfo
  {author} {\bibfnamefont {K.}~\bibnamefont {Helmerson}}, \ and\ \bibinfo
  {author} {\bibfnamefont {W.~D.}\ \bibnamefont {Phillips}},\ }\bibfield
  {title} {\enquote {\bibinfo {title} {{Quantized Rotation of Atoms from
  Photons with Orbital Angular Momentum}},}\ }\href {\doibase
  10.1103/PhysRevLett.97.170406} {\bibfield  {journal} {\bibinfo  {journal}
  {Phys. Rev. Lett.}\ }\textbf {\bibinfo {volume} {97}},\ \bibinfo {pages}
  {170406} (\bibinfo {year} {2006})}\BibitemShut {NoStop}%
\bibitem [{\citenamefont {Kwon}\ \emph {et~al.}(2019)\citenamefont {Kwon},
  \citenamefont {Oh}, \citenamefont {Gong}, \citenamefont {Kim}, \citenamefont
  {Kang}, \citenamefont {Kang}, \citenamefont {Song}, \citenamefont {Choi},\
  and\ \citenamefont {Cho}}]{kwon19}%
  \BibitemOpen
  \bibfield  {author} {\bibinfo {author} {\bibfnamefont {Min-Sik}\ \bibnamefont
  {Kwon}}, \bibinfo {author} {\bibfnamefont {Byoung~Yong}\ \bibnamefont {Oh}},
  \bibinfo {author} {\bibfnamefont {Su-Hyun}\ \bibnamefont {Gong}}, \bibinfo
  {author} {\bibfnamefont {Je-Hyung}\ \bibnamefont {Kim}}, \bibinfo {author}
  {\bibfnamefont {Hang~Kyu}\ \bibnamefont {Kang}}, \bibinfo {author}
  {\bibfnamefont {Sooseok}\ \bibnamefont {Kang}}, \bibinfo {author}
  {\bibfnamefont {Jin~Dong}\ \bibnamefont {Song}}, \bibinfo {author}
  {\bibfnamefont {Hyoungsoon}\ \bibnamefont {Choi}}, \ and\ \bibinfo {author}
  {\bibfnamefont {Yong-Hoon}\ \bibnamefont {Cho}},\ }\bibfield  {title}
  {\enquote {\bibinfo {title} {{Direct Transfer of Light's Orbital Angular
  Momentum onto a Nonresonantly Excited Polariton Superfluid}},}\ }\href
  {\doibase 10.1103/PhysRevLett.122.045302} {\bibfield  {journal} {\bibinfo
  {journal} {Phys. Rev. Lett.}\ }\textbf {\bibinfo {volume} {122}},\ \bibinfo
  {pages} {045302} (\bibinfo {year} {2019})}\BibitemShut {NoStop}%
\bibitem [{\citenamefont {Shen}\ \emph {et~al.}(2019)\citenamefont {Shen},
  \citenamefont {Wang}, \citenamefont {Xie}, \citenamefont {Min}, \citenamefont
  {Fu}, \citenamefont {Liu}, \citenamefont {Gong},\ and\ \citenamefont
  {Yuan}}]{shen19}%
  \BibitemOpen
  \bibfield  {author} {\bibinfo {author} {\bibfnamefont {Yijie}\ \bibnamefont
  {Shen}}, \bibinfo {author} {\bibfnamefont {Xuejiao}\ \bibnamefont {Wang}},
  \bibinfo {author} {\bibfnamefont {Zhenwei}\ \bibnamefont {Xie}}, \bibinfo
  {author} {\bibfnamefont {Changjun}\ \bibnamefont {Min}}, \bibinfo {author}
  {\bibfnamefont {Xing}\ \bibnamefont {Fu}}, \bibinfo {author} {\bibfnamefont
  {Qiang}\ \bibnamefont {Liu}}, \bibinfo {author} {\bibfnamefont {Mali}\
  \bibnamefont {Gong}}, \ and\ \bibinfo {author} {\bibfnamefont {Xiaocong}\
  \bibnamefont {Yuan}},\ }\bibfield  {title} {\enquote {\bibinfo {title}
  {{Optical vortices 30 years on: OAM manipulation from topological charge to
  multiple singularities}},}\ }\href {\doibase 10.1038/s41377-019-0194-2}
  {\bibfield  {journal} {\bibinfo  {journal} {Light: Science \& Applications}\
  }\textbf {\bibinfo {volume} {8}},\ \bibinfo {pages} {90} (\bibinfo {year}
  {2019})}\BibitemShut {NoStop}%
\bibitem [{\citenamefont {Wang}\ \emph {et~al.}(2012)\citenamefont {Wang},
  \citenamefont {Yang}, \citenamefont {Fazal}, \citenamefont {Ahmed},
  \citenamefont {Yan}, \citenamefont {Huang}, \citenamefont {Ren},
  \citenamefont {Yue}, \citenamefont {Dolinar}, \citenamefont {Tur},\ and\
  \citenamefont {Willner}}]{wang12}%
  \BibitemOpen
  \bibfield  {author} {\bibinfo {author} {\bibfnamefont {Jian}\ \bibnamefont
  {Wang}}, \bibinfo {author} {\bibfnamefont {Jeng-Yuan}\ \bibnamefont {Yang}},
  \bibinfo {author} {\bibfnamefont {Irfan~M.}\ \bibnamefont {Fazal}}, \bibinfo
  {author} {\bibfnamefont {Nisar}\ \bibnamefont {Ahmed}}, \bibinfo {author}
  {\bibfnamefont {Yan}\ \bibnamefont {Yan}}, \bibinfo {author} {\bibfnamefont
  {Hao}\ \bibnamefont {Huang}}, \bibinfo {author} {\bibfnamefont {Yongxiong}\
  \bibnamefont {Ren}}, \bibinfo {author} {\bibfnamefont {Yang}\ \bibnamefont
  {Yue}}, \bibinfo {author} {\bibfnamefont {Samuel}\ \bibnamefont {Dolinar}},
  \bibinfo {author} {\bibfnamefont {Moshe}\ \bibnamefont {Tur}}, \ and\
  \bibinfo {author} {\bibfnamefont {Alan~E.}\ \bibnamefont {Willner}},\
  }\bibfield  {title} {\enquote {\bibinfo {title} {{Terabit free-space data
  transmission employing orbital angular momentum multiplexing}},}\ }\href
  {\doibase 10.1038/nphoton.2012.138} {\bibfield  {journal} {\bibinfo
  {journal} {Nature Photonics}\ }\textbf {\bibinfo {volume} {6}},\ \bibinfo
  {pages} {488--496} (\bibinfo {year} {2012})}\BibitemShut {NoStop}%
\bibitem [{\citenamefont {Bozinovic}\ \emph {et~al.}(2013)\citenamefont
  {Bozinovic}, \citenamefont {Yue}, \citenamefont {Ren}, \citenamefont {Tur},
  \citenamefont {Kristensen}, \citenamefont {Huang}, \citenamefont {Willner},\
  and\ \citenamefont {Ramachandran}}]{bozinovic13}%
  \BibitemOpen
  \bibfield  {author} {\bibinfo {author} {\bibfnamefont {Nenad}\ \bibnamefont
  {Bozinovic}}, \bibinfo {author} {\bibfnamefont {Yang}\ \bibnamefont {Yue}},
  \bibinfo {author} {\bibfnamefont {Yongxiong}\ \bibnamefont {Ren}}, \bibinfo
  {author} {\bibfnamefont {Moshe}\ \bibnamefont {Tur}}, \bibinfo {author}
  {\bibfnamefont {Poul}\ \bibnamefont {Kristensen}}, \bibinfo {author}
  {\bibfnamefont {Hao}\ \bibnamefont {Huang}}, \bibinfo {author} {\bibfnamefont
  {Alan~E.}\ \bibnamefont {Willner}}, \ and\ \bibinfo {author} {\bibfnamefont
  {Siddharth}\ \bibnamefont {Ramachandran}},\ }\bibfield  {title} {\enquote
  {\bibinfo {title} {{Terabit-Scale Orbital Angular Momentum Mode Division
  Multiplexing in Fibers}},}\ }\href
  {https://science.sciencemag.org/content/340/6140/1545} {\bibfield  {journal}
  {\bibinfo  {journal} {Science}\ }\textbf {\bibinfo {volume} {340}},\ \bibinfo
  {pages} {1545--1548} (\bibinfo {year} {2013})}\BibitemShut {NoStop}%
\bibitem [{\citenamefont {Nicolas}\ \emph {et~al.}(2014)\citenamefont
  {Nicolas}, \citenamefont {Veissier}, \citenamefont {Giner}, \citenamefont
  {Giacobino}, \citenamefont {Maxein},\ and\ \citenamefont
  {Laurat}}]{nicolas14}%
  \BibitemOpen
  \bibfield  {author} {\bibinfo {author} {\bibfnamefont {A.}~\bibnamefont
  {Nicolas}}, \bibinfo {author} {\bibfnamefont {L.}~\bibnamefont {Veissier}},
  \bibinfo {author} {\bibfnamefont {L.}~\bibnamefont {Giner}}, \bibinfo
  {author} {\bibfnamefont {E.}~\bibnamefont {Giacobino}}, \bibinfo {author}
  {\bibfnamefont {D.}~\bibnamefont {Maxein}}, \ and\ \bibinfo {author}
  {\bibfnamefont {J.}~\bibnamefont {Laurat}},\ }\bibfield  {title} {\enquote
  {\bibinfo {title} {{A quantum memory for orbital angular momentum photonic
  qubits}},}\ }\href {\doibase 10.1038/nphoton.2013.355} {\bibfield  {journal}
  {\bibinfo  {journal} {Nature Photonics}\ }\textbf {\bibinfo {volume} {8}},\
  \bibinfo {pages} {234--238} (\bibinfo {year} {2014})}\BibitemShut {NoStop}%
\bibitem [{\citenamefont {Tamburini}\ \emph {et~al.}(2011)\citenamefont
  {Tamburini}, \citenamefont {Thid{\'e}}, \citenamefont {Molina-Terriza},\ and\
  \citenamefont {Anzolin}}]{tamburini11}%
  \BibitemOpen
  \bibfield  {author} {\bibinfo {author} {\bibfnamefont {Fabrizio}\
  \bibnamefont {Tamburini}}, \bibinfo {author} {\bibfnamefont {Bo}~\bibnamefont
  {Thid{\'e}}}, \bibinfo {author} {\bibfnamefont {Gabriel}\ \bibnamefont
  {Molina-Terriza}}, \ and\ \bibinfo {author} {\bibfnamefont {Gabriele}\
  \bibnamefont {Anzolin}},\ }\bibfield  {title} {\enquote {\bibinfo {title}
  {{Twisting of light around rotating black holes}},}\ }\href {\doibase
  10.1038/nphys1907} {\bibfield  {journal} {\bibinfo  {journal} {Nature
  Physics}\ }\textbf {\bibinfo {volume} {7}},\ \bibinfo {pages} {195--197}
  (\bibinfo {year} {2011})}\BibitemShut {NoStop}%
\bibitem [{\citenamefont {Kong}\ \emph {et~al.}(2017)\citenamefont {Kong},
  \citenamefont {Zhang}, \citenamefont {Bouchard}, \citenamefont {Li},
  \citenamefont {Brown}, \citenamefont {Ko}, \citenamefont {Hammond},
  \citenamefont {Arissian}, \citenamefont {Boyd}, \citenamefont {Karimi},\ and\
  \citenamefont {Corkum}}]{kong17}%
  \BibitemOpen
  \bibfield  {author} {\bibinfo {author} {\bibfnamefont {Fanqi}\ \bibnamefont
  {Kong}}, \bibinfo {author} {\bibfnamefont {Chunmei}\ \bibnamefont {Zhang}},
  \bibinfo {author} {\bibfnamefont {Fr{\'e}d{\'e}ric}\ \bibnamefont
  {Bouchard}}, \bibinfo {author} {\bibfnamefont {Zhengyan}\ \bibnamefont {Li}},
  \bibinfo {author} {\bibfnamefont {Graham~G.}\ \bibnamefont {Brown}}, \bibinfo
  {author} {\bibfnamefont {Dong~Hyuk}\ \bibnamefont {Ko}}, \bibinfo {author}
  {\bibfnamefont {T.~J.}\ \bibnamefont {Hammond}}, \bibinfo {author}
  {\bibfnamefont {Ladan}\ \bibnamefont {Arissian}}, \bibinfo {author}
  {\bibfnamefont {Robert~W.}\ \bibnamefont {Boyd}}, \bibinfo {author}
  {\bibfnamefont {Ebrahim}\ \bibnamefont {Karimi}}, \ and\ \bibinfo {author}
  {\bibfnamefont {P.~B.}\ \bibnamefont {Corkum}},\ }\bibfield  {title}
  {\enquote {\bibinfo {title} {{Controlling the orbital angular momentum of
  high harmonic vortices}},}\ }\href {\doibase 10.1038/ncomms14970} {\bibfield
  {journal} {\bibinfo  {journal} {Nature Communications}\ }\textbf {\bibinfo
  {volume} {8}},\ \bibinfo {pages} {14970} (\bibinfo {year}
  {2017})}\BibitemShut {NoStop}%
\bibitem [{\citenamefont {Gauthier}\ \emph {et~al.}(2017)\citenamefont
  {Gauthier}, \citenamefont {Ribi\v{c}}, \citenamefont {Adhikary},
  \citenamefont {Camper}, \citenamefont {Chappuis}, \citenamefont {Cucini},
  \citenamefont {DiMauro}, \citenamefont {Dovillaire}, \citenamefont
  {Frassetto}, \citenamefont {G{\'e}neaux}, \citenamefont {Miotti},
  \citenamefont {Poletto}, \citenamefont {Ressel}, \citenamefont {Spezzani},
  \citenamefont {Stupar}, \citenamefont {Ruchon},\ and\ \citenamefont {{De
  Ninno}}}]{gauthier17}%
  \BibitemOpen
  \bibfield  {author} {\bibinfo {author} {\bibfnamefont {D.}~\bibnamefont
  {Gauthier}}, \bibinfo {author} {\bibfnamefont {P.~Rebernik}\ \bibnamefont
  {Ribi\v{c}}}, \bibinfo {author} {\bibfnamefont {G.}~\bibnamefont {Adhikary}},
  \bibinfo {author} {\bibfnamefont {A.}~\bibnamefont {Camper}}, \bibinfo
  {author} {\bibfnamefont {C.}~\bibnamefont {Chappuis}}, \bibinfo {author}
  {\bibfnamefont {R.}~\bibnamefont {Cucini}}, \bibinfo {author} {\bibfnamefont
  {L.~F.}\ \bibnamefont {DiMauro}}, \bibinfo {author} {\bibfnamefont
  {G.}~\bibnamefont {Dovillaire}}, \bibinfo {author} {\bibfnamefont
  {F.}~\bibnamefont {Frassetto}}, \bibinfo {author} {\bibfnamefont
  {R.}~\bibnamefont {G{\'e}neaux}}, \bibinfo {author} {\bibfnamefont
  {P.}~\bibnamefont {Miotti}}, \bibinfo {author} {\bibfnamefont
  {L.}~\bibnamefont {Poletto}}, \bibinfo {author} {\bibfnamefont
  {B.}~\bibnamefont {Ressel}}, \bibinfo {author} {\bibfnamefont
  {C.}~\bibnamefont {Spezzani}}, \bibinfo {author} {\bibfnamefont
  {M.}~\bibnamefont {Stupar}}, \bibinfo {author} {\bibfnamefont
  {T.}~\bibnamefont {Ruchon}}, \ and\ \bibinfo {author} {\bibfnamefont
  {G.}~\bibnamefont {{De Ninno}}},\ }\bibfield  {title} {\enquote {\bibinfo
  {title} {{Tunable orbital angular momentum in high-harmonic generation}},}\
  }\href {\doibase 10.1038/ncomms14971} {\bibfield  {journal} {\bibinfo
  {journal} {Nature Communications}\ }\textbf {\bibinfo {volume} {8}},\
  \bibinfo {pages} {14971} (\bibinfo {year} {2017})}\BibitemShut {NoStop}%
\bibitem [{\citenamefont {Haldane}(1988)}]{PhysRevLett.61.2015}%
  \BibitemOpen
  \bibfield  {author} {\bibinfo {author} {\bibfnamefont {F.~D.~M.}\
  \bibnamefont {Haldane}},\ }\bibfield  {title} {\enquote {\bibinfo {title}
  {{Model for a Quantum Hall Effect without Landau Levels: Condensed-Matter
  Realization of the "Parity Anomaly"}},}\ }\href {\doibase
  10.1103/PhysRevLett.61.2015} {\bibfield  {journal} {\bibinfo  {journal}
  {Phys. Rev. Lett.}\ }\textbf {\bibinfo {volume} {61}},\ \bibinfo {pages}
  {2015--2018} (\bibinfo {year} {1988})}\BibitemShut {NoStop}%
\bibitem [{\citenamefont {Chang}\ \emph {et~al.}(2013)\citenamefont {Chang},
  \citenamefont {Zhang}, \citenamefont {Feng}, \citenamefont {Shen},
  \citenamefont {Zhang}, \citenamefont {Guo}, \citenamefont {Li}, \citenamefont
  {Ou}, \citenamefont {Wei}, \citenamefont {Wang}, \citenamefont {Ji},
  \citenamefont {Feng}, \citenamefont {Ji}, \citenamefont {Chen}, \citenamefont
  {Jia}, \citenamefont {Dai}, \citenamefont {Fang}, \citenamefont {Zhang},
  \citenamefont {He}, \citenamefont {Wang}, \citenamefont {Lu}, \citenamefont
  {Ma},\ and\ \citenamefont {Xue}}]{chang13}%
  \BibitemOpen
  \bibfield  {author} {\bibinfo {author} {\bibfnamefont {Cui-Zu}\ \bibnamefont
  {Chang}}, \bibinfo {author} {\bibfnamefont {Jinsong}\ \bibnamefont {Zhang}},
  \bibinfo {author} {\bibfnamefont {Xiao}\ \bibnamefont {Feng}}, \bibinfo
  {author} {\bibfnamefont {Jie}\ \bibnamefont {Shen}}, \bibinfo {author}
  {\bibfnamefont {Zuocheng}\ \bibnamefont {Zhang}}, \bibinfo {author}
  {\bibfnamefont {Minghua}\ \bibnamefont {Guo}}, \bibinfo {author}
  {\bibfnamefont {Kang}\ \bibnamefont {Li}}, \bibinfo {author} {\bibfnamefont
  {Yunbo}\ \bibnamefont {Ou}}, \bibinfo {author} {\bibfnamefont {Pang}\
  \bibnamefont {Wei}}, \bibinfo {author} {\bibfnamefont {Li-Li}\ \bibnamefont
  {Wang}}, \bibinfo {author} {\bibfnamefont {Zhong-Qing}\ \bibnamefont {Ji}},
  \bibinfo {author} {\bibfnamefont {Yang}\ \bibnamefont {Feng}}, \bibinfo
  {author} {\bibfnamefont {Shuaihua}\ \bibnamefont {Ji}}, \bibinfo {author}
  {\bibfnamefont {Xi}~\bibnamefont {Chen}}, \bibinfo {author} {\bibfnamefont
  {Jinfeng}\ \bibnamefont {Jia}}, \bibinfo {author} {\bibfnamefont
  {Xi}~\bibnamefont {Dai}}, \bibinfo {author} {\bibfnamefont {Zhong}\
  \bibnamefont {Fang}}, \bibinfo {author} {\bibfnamefont {Shou-Cheng}\
  \bibnamefont {Zhang}}, \bibinfo {author} {\bibfnamefont {Ke}~\bibnamefont
  {He}}, \bibinfo {author} {\bibfnamefont {Yayu}\ \bibnamefont {Wang}},
  \bibinfo {author} {\bibfnamefont {Li}~\bibnamefont {Lu}}, \bibinfo {author}
  {\bibfnamefont {Xu-Cun}\ \bibnamefont {Ma}}, \ and\ \bibinfo {author}
  {\bibfnamefont {Qi-Kun}\ \bibnamefont {Xue}},\ }\bibfield  {title} {\enquote
  {\bibinfo {title} {{Experimental Observation of the Quantum Anomalous Hall
  Effect in a Magnetic Topological Insulator}},}\ }\href
  {https://science.sciencemag.org/content/340/6129/167} {\bibfield  {journal}
  {\bibinfo  {journal} {Science}\ }\textbf {\bibinfo {volume} {340}},\ \bibinfo
  {pages} {167--170} (\bibinfo {year} {2013})}\BibitemShut {NoStop}%
\bibitem [{\citenamefont {Rechtsman}\ \emph {et~al.}(2013)\citenamefont
  {Rechtsman}, \citenamefont {Zeuner}, \citenamefont {Plotnik}, \citenamefont
  {Lumer}, \citenamefont {Podolsky}, \citenamefont {Dreisow}, \citenamefont
  {Nolte}, \citenamefont {Segev},\ and\ \citenamefont {Szameit}}]{rechtsman13}%
  \BibitemOpen
  \bibfield  {author} {\bibinfo {author} {\bibfnamefont {Mikael~C.}\
  \bibnamefont {Rechtsman}}, \bibinfo {author} {\bibfnamefont {Julia~M.}\
  \bibnamefont {Zeuner}}, \bibinfo {author} {\bibfnamefont {Yonatan}\
  \bibnamefont {Plotnik}}, \bibinfo {author} {\bibfnamefont {Yaakov}\
  \bibnamefont {Lumer}}, \bibinfo {author} {\bibfnamefont {Daniel}\
  \bibnamefont {Podolsky}}, \bibinfo {author} {\bibfnamefont {Felix}\
  \bibnamefont {Dreisow}}, \bibinfo {author} {\bibfnamefont {Stefan}\
  \bibnamefont {Nolte}}, \bibinfo {author} {\bibfnamefont {Mordechai}\
  \bibnamefont {Segev}}, \ and\ \bibinfo {author} {\bibfnamefont {Alexander}\
  \bibnamefont {Szameit}},\ }\bibfield  {title} {\enquote {\bibinfo {title}
  {{Photonic Floquet topological insulators}},}\ }\href {\doibase
  10.1038/nature12066} {\bibfield  {journal} {\bibinfo  {journal} {Nature}\
  }\textbf {\bibinfo {volume} {496}},\ \bibinfo {pages} {196--200} (\bibinfo
  {year} {2013})}\BibitemShut {NoStop}%
\bibitem [{\citenamefont {Jotzu}\ \emph {et~al.}(2014)\citenamefont {Jotzu},
  \citenamefont {Messer}, \citenamefont {Desbuquois}, \citenamefont {Lebrat},
  \citenamefont {Uehlinger}, \citenamefont {Greif},\ and\ \citenamefont
  {Esslinger}}]{jotzu2014experimental}%
  \BibitemOpen
  \bibfield  {author} {\bibinfo {author} {\bibfnamefont {Gregor}\ \bibnamefont
  {Jotzu}}, \bibinfo {author} {\bibfnamefont {Michael}\ \bibnamefont {Messer}},
  \bibinfo {author} {\bibfnamefont {R{\'e}mi}\ \bibnamefont {Desbuquois}},
  \bibinfo {author} {\bibfnamefont {Martin}\ \bibnamefont {Lebrat}}, \bibinfo
  {author} {\bibfnamefont {Thomas}\ \bibnamefont {Uehlinger}}, \bibinfo
  {author} {\bibfnamefont {Daniel}\ \bibnamefont {Greif}}, \ and\ \bibinfo
  {author} {\bibfnamefont {Tilman}\ \bibnamefont {Esslinger}},\ }\bibfield
  {title} {\enquote {\bibinfo {title} {{Experimental realization of the
  topological Haldane model with ultracold fermions}},}\ }\href
  {https://www.nature.com/articles/nature13915} {\bibfield  {journal} {\bibinfo
   {journal} {Nature}\ }\textbf {\bibinfo {volume} {515}},\ \bibinfo {pages}
  {237--240} (\bibinfo {year} {2014})}\BibitemShut {NoStop}%
\bibitem [{\citenamefont {McIver}\ \emph {et~al.}(2020)\citenamefont {McIver},
  \citenamefont {Schulte}, \citenamefont {Stein}, \citenamefont {Matsuyama},
  \citenamefont {Jotzu}, \citenamefont {Meier},\ and\ \citenamefont
  {Cavalleri}}]{mciver20}%
  \BibitemOpen
  \bibfield  {author} {\bibinfo {author} {\bibfnamefont {J.~W.}\ \bibnamefont
  {McIver}}, \bibinfo {author} {\bibfnamefont {B.}~\bibnamefont {Schulte}},
  \bibinfo {author} {\bibfnamefont {F.-U.}\ \bibnamefont {Stein}}, \bibinfo
  {author} {\bibfnamefont {T.}~\bibnamefont {Matsuyama}}, \bibinfo {author}
  {\bibfnamefont {G.}~\bibnamefont {Jotzu}}, \bibinfo {author} {\bibfnamefont
  {G.}~\bibnamefont {Meier}}, \ and\ \bibinfo {author} {\bibfnamefont
  {A.}~\bibnamefont {Cavalleri}},\ }\bibfield  {title} {\enquote {\bibinfo
  {title} {{Light-induced anomalous Hall effect in graphene}},}\ }\href
  {\doibase 10.1038/s41567-019-0698-y} {\bibfield  {journal} {\bibinfo
  {journal} {Nature Physics}\ }\textbf {\bibinfo {volume} {16}},\ \bibinfo
  {pages} {38--41} (\bibinfo {year} {2020})}\BibitemShut {NoStop}%
\bibitem [{\citenamefont {Oka}\ and\ \citenamefont {Aoki}(2009)}]{oka2009}%
  \BibitemOpen
  \bibfield  {author} {\bibinfo {author} {\bibfnamefont {Takashi}\ \bibnamefont
  {Oka}}\ and\ \bibinfo {author} {\bibfnamefont {Hideo}\ \bibnamefont {Aoki}},\
  }\bibfield  {title} {\enquote {\bibinfo {title} {{Photovoltaic Hall effect in
  graphene}},}\ }\href {https://aip.scitation.org/doi/full/10.1063/1.5027667}
  {\bibfield  {journal} {\bibinfo  {journal} {Phys. Rev. B}\ }\textbf {\bibinfo
  {volume} {79}},\ \bibinfo {pages} {081406(R)} (\bibinfo {year}
  {2009})}\BibitemShut {NoStop}%
\bibitem [{\citenamefont {{Castro Neto}}\ \emph {et~al.}(2009)\citenamefont
  {{Castro Neto}}, \citenamefont {Guinea}, \citenamefont {Peres}, \citenamefont
  {Novoselov},\ and\ \citenamefont {Geim}}]{castro_neto09}%
  \BibitemOpen
  \bibfield  {author} {\bibinfo {author} {\bibfnamefont {A.~H.}\ \bibnamefont
  {{Castro Neto}}}, \bibinfo {author} {\bibfnamefont {F.}~\bibnamefont
  {Guinea}}, \bibinfo {author} {\bibfnamefont {N.~M.~R.}\ \bibnamefont
  {Peres}}, \bibinfo {author} {\bibfnamefont {K.~S.}\ \bibnamefont
  {Novoselov}}, \ and\ \bibinfo {author} {\bibfnamefont {A.~K.}\ \bibnamefont
  {Geim}},\ }\bibfield  {title} {\enquote {\bibinfo {title} {{The electronic
  properties of graphene}},}\ }\href {\doibase 10.1103/RevModPhys.81.109}
  {\bibfield  {journal} {\bibinfo  {journal} {Rev. Mod. Phys.}\ }\textbf
  {\bibinfo {volume} {81}},\ \bibinfo {pages} {109--162} (\bibinfo {year}
  {2009})}\BibitemShut {NoStop}%
\bibitem [{Note1()}]{Note1}%
  \BibitemOpen
  \bibinfo {note} {\label {SM} See Supplemental Material which describes the
  used methods, and which contains reference to Ref. \cite
  {Eckardt_2015}.}\BibitemShut {Stop}%
\bibitem [{\citenamefont {Zhou}\ \emph {et~al.}(2007)\citenamefont {Zhou},
  \citenamefont {Gweon}, \citenamefont {Fedorov}, \citenamefont {First},
  \citenamefont {de~Heer}, \citenamefont {Lee}, \citenamefont {Guinea},
  \citenamefont {{Castro Neto}},\ and\ \citenamefont {Lanzara}}]{zhou07}%
  \BibitemOpen
  \bibfield  {author} {\bibinfo {author} {\bibfnamefont {S.~Y.}\ \bibnamefont
  {Zhou}}, \bibinfo {author} {\bibfnamefont {G.-H.}\ \bibnamefont {Gweon}},
  \bibinfo {author} {\bibfnamefont {A.~V.}\ \bibnamefont {Fedorov}}, \bibinfo
  {author} {\bibfnamefont {P.~N.}\ \bibnamefont {First}}, \bibinfo {author}
  {\bibfnamefont {W.~A.}\ \bibnamefont {de~Heer}}, \bibinfo {author}
  {\bibfnamefont {D.-H.}\ \bibnamefont {Lee}}, \bibinfo {author} {\bibfnamefont
  {F.}~\bibnamefont {Guinea}}, \bibinfo {author} {\bibfnamefont {A.~H.}\
  \bibnamefont {{Castro Neto}}}, \ and\ \bibinfo {author} {\bibfnamefont
  {A.}~\bibnamefont {Lanzara}},\ }\bibfield  {title} {\enquote {\bibinfo
  {title} {{Substrate-induced bandgap opening in epitaxial graphene}},}\ }\href
  {\doibase 10.1038/nmat2003} {\bibfield  {journal} {\bibinfo  {journal}
  {Nature Materials}\ }\textbf {\bibinfo {volume} {6}},\ \bibinfo {pages}
  {770--775} (\bibinfo {year} {2007})}\BibitemShut {NoStop}%
\bibitem [{\citenamefont {Semenoff}(1984)}]{semenoff84}%
  \BibitemOpen
  \bibfield  {author} {\bibinfo {author} {\bibfnamefont {Gordon~W.}\
  \bibnamefont {Semenoff}},\ }\bibfield  {title} {\enquote {\bibinfo {title}
  {{Condensed-Matter Simulation of a Three-Dimensional Anomaly}},}\ }\href
  {\doibase 10.1103/PhysRevLett.53.2449} {\bibfield  {journal} {\bibinfo
  {journal} {Phys. Rev. Lett.}\ }\textbf {\bibinfo {volume} {53}},\ \bibinfo
  {pages} {2449--2452} (\bibinfo {year} {1984})}\BibitemShut {NoStop}%
\bibitem [{\citenamefont {Bukov}\ \emph {et~al.}(2015)\citenamefont {Bukov},
  \citenamefont {D'Alessio},\ and\ \citenamefont {Polkovnikov}}]{bukov15}%
  \BibitemOpen
  \bibfield  {author} {\bibinfo {author} {\bibfnamefont {Marin}\ \bibnamefont
  {Bukov}}, \bibinfo {author} {\bibfnamefont {Luca}\ \bibnamefont {D'Alessio}},
  \ and\ \bibinfo {author} {\bibfnamefont {Anatoli}\ \bibnamefont
  {Polkovnikov}},\ }\bibfield  {title} {\enquote {\bibinfo {title} {{Universal
  high-frequency behavior of periodically driven systems: from dynamical
  stabilization to Floquet engineering}},}\ }\href {\doibase
  10.1080/00018732.2015.1055918} {\bibfield  {journal} {\bibinfo  {journal}
  {Advances in Physics}\ }\textbf {\bibinfo {volume} {64}},\ \bibinfo {pages}
  {139--226} (\bibinfo {year} {2015})}\BibitemShut {NoStop}%
\bibitem [{\citenamefont {Bianco}\ and\ \citenamefont
  {Resta}(2011)}]{PhysRevB.84.241106}%
  \BibitemOpen
  \bibfield  {author} {\bibinfo {author} {\bibfnamefont {Raffaello}\
  \bibnamefont {Bianco}}\ and\ \bibinfo {author} {\bibfnamefont {Raffaele}\
  \bibnamefont {Resta}},\ }\bibfield  {title} {\enquote {\bibinfo {title}
  {{Mapping topological order in coordinate space}},}\ }\href {\doibase
  10.1103/PhysRevB.84.241106} {\bibfield  {journal} {\bibinfo  {journal} {Phys.
  Rev. B}\ }\textbf {\bibinfo {volume} {84}},\ \bibinfo {pages} {241106}
  (\bibinfo {year} {2011})}\BibitemShut {NoStop}%
\bibitem [{\citenamefont {He}\ \emph {et~al.}(2019)\citenamefont {He},
  \citenamefont {Ding}, \citenamefont {Zhou}, \citenamefont {Wang},\ and\
  \citenamefont {Gong}}]{PhysRevB.100.214109}%
  \BibitemOpen
  \bibfield  {author} {\bibinfo {author} {\bibfnamefont {Ai-Lei}\ \bibnamefont
  {He}}, \bibinfo {author} {\bibfnamefont {Lu-Rong}\ \bibnamefont {Ding}},
  \bibinfo {author} {\bibfnamefont {Yuan}\ \bibnamefont {Zhou}}, \bibinfo
  {author} {\bibfnamefont {Yi-Fei}\ \bibnamefont {Wang}}, \ and\ \bibinfo
  {author} {\bibfnamefont {Chang-De}\ \bibnamefont {Gong}},\ }\bibfield
  {title} {\enquote {\bibinfo {title} {{Quasicrystalline Chern insulators}},}\
  }\href {\doibase 10.1103/PhysRevB.100.214109} {\bibfield  {journal} {\bibinfo
   {journal} {Phys. Rev. B}\ }\textbf {\bibinfo {volume} {100}},\ \bibinfo
  {pages} {214109} (\bibinfo {year} {2019})}\BibitemShut {NoStop}%
\bibitem [{\citenamefont {T{\"o}r{\"o}k}\ and\ \citenamefont
  {Munro}(2004)}]{Torok2004}%
  \BibitemOpen
  \bibfield  {author} {\bibinfo {author} {\bibfnamefont {P.}~\bibnamefont
  {T{\"o}r{\"o}k}}\ and\ \bibinfo {author} {\bibfnamefont {P.R.T.}\
  \bibnamefont {Munro}},\ }\bibfield  {title} {\enquote {\bibinfo {title} {{The
  use of Gauss-Laguerre vector beams in STED microscopy}},}\ }\href
  {https://www.osapublishing.org/abstract.cfm?URI=oe-12-15-3605} {\bibfield
  {journal} {\bibinfo  {journal} {Optics Express}\ }\textbf {\bibinfo {volume}
  {12}},\ \bibinfo {pages} {3605--3617} (\bibinfo {year} {2004})}\BibitemShut
  {NoStop}%
\bibitem [{\citenamefont {Polini}\ \emph {et~al.}(2013)\citenamefont {Polini},
  \citenamefont {Guinea}, \citenamefont {Lewenstein}, \citenamefont
  {Manoharan},\ and\ \citenamefont {Pellegrini}}]{polini2013artificial}%
  \BibitemOpen
  \bibfield  {author} {\bibinfo {author} {\bibfnamefont {Marco}\ \bibnamefont
  {Polini}}, \bibinfo {author} {\bibfnamefont {Francisco}\ \bibnamefont
  {Guinea}}, \bibinfo {author} {\bibfnamefont {Maciej}\ \bibnamefont
  {Lewenstein}}, \bibinfo {author} {\bibfnamefont {Hari~C}\ \bibnamefont
  {Manoharan}}, \ and\ \bibinfo {author} {\bibfnamefont {Vittorio}\
  \bibnamefont {Pellegrini}},\ }\bibfield  {title} {\enquote {\bibinfo {title}
  {{Artificial honeycomb lattices for electrons, atoms and photons}},}\
  }\href@noop {} {\bibfield  {journal} {\bibinfo  {journal} {Nature
  nanotechnology}\ }\textbf {\bibinfo {volume} {8}},\ \bibinfo {pages} {625}
  (\bibinfo {year} {2013})}\BibitemShut {NoStop}%
\bibitem [{\citenamefont {Klar}\ \emph {et~al.}(2001)\citenamefont {Klar},
  \citenamefont {Engel},\ and\ \citenamefont {Hell}}]{Klar2001}%
  \BibitemOpen
  \bibfield  {author} {\bibinfo {author} {\bibfnamefont {Thomas~A.}\
  \bibnamefont {Klar}}, \bibinfo {author} {\bibfnamefont {Egbert}\ \bibnamefont
  {Engel}}, \ and\ \bibinfo {author} {\bibfnamefont {Stefan~W.}\ \bibnamefont
  {Hell}},\ }\bibfield  {title} {\enquote {\bibinfo {title} {{Breaking
  Abbe{\rq}s diffraction resolution limit in fluorescence microscopy with
  stimulated emission depletion beams of various shapes}},}\ }\href
  {https://link.aps.org/doi/10.1103/PhysRevE.64.066613} {\bibfield  {journal}
  {\bibinfo  {journal} {Physical Review E}\ }\textbf {\bibinfo {volume} {64}},\
  \bibinfo {pages} {066613} (\bibinfo {year} {2001})}\BibitemShut {NoStop}%
\bibitem [{\citenamefont {Pisanty}\ \emph
  {et~al.}(2019{\natexlab{a}})\citenamefont {Pisanty}, \citenamefont {Machado},
  \citenamefont {Vicu{\~n}a-Hern{\'a}ndez}, \citenamefont {Pic{\'o}n},
  \citenamefont {Celi}, \citenamefont {Torres},\ and\ \citenamefont
  {Lewenstein}}]{pisanty2019knotting}%
  \BibitemOpen
  \bibfield  {author} {\bibinfo {author} {\bibfnamefont {Emilio}\ \bibnamefont
  {Pisanty}}, \bibinfo {author} {\bibfnamefont {Gerard~J}\ \bibnamefont
  {Machado}}, \bibinfo {author} {\bibfnamefont {Ver{\'o}nica}\ \bibnamefont
  {Vicu{\~n}a-Hern{\'a}ndez}}, \bibinfo {author} {\bibfnamefont {Antonio}\
  \bibnamefont {Pic{\'o}n}}, \bibinfo {author} {\bibfnamefont {Alessio}\
  \bibnamefont {Celi}}, \bibinfo {author} {\bibfnamefont {Juan~P}\ \bibnamefont
  {Torres}}, \ and\ \bibinfo {author} {\bibfnamefont {Maciej}\ \bibnamefont
  {Lewenstein}},\ }\bibfield  {title} {\enquote {\bibinfo {title} {{Knotting
  fractional-order knots with the polarization state of light}},}\ }\href@noop
  {} {\bibfield  {journal} {\bibinfo  {journal} {Nature Photonics}\ }\textbf
  {\bibinfo {volume} {13}},\ \bibinfo {pages} {569--574} (\bibinfo {year}
  {2019}{\natexlab{a}})}\BibitemShut {NoStop}%
\bibitem [{\citenamefont {Pisanty}\ \emph
  {et~al.}(2019{\natexlab{b}})\citenamefont {Pisanty}, \citenamefont {Rego},
  \citenamefont {{San Rom{\'a}n}}, \citenamefont {Pic{\'o}n}, \citenamefont
  {Dorney}, \citenamefont {Kapteyn}, \citenamefont {Murnane}, \citenamefont
  {Plaja}, \citenamefont {Lewenstein},\ and\ \citenamefont
  {Hern{\'a}ndez-Garc{\'i}a}}]{PhysRevLett.122.203201}%
  \BibitemOpen
  \bibfield  {author} {\bibinfo {author} {\bibfnamefont {Emilio}\ \bibnamefont
  {Pisanty}}, \bibinfo {author} {\bibfnamefont {Laura}\ \bibnamefont {Rego}},
  \bibinfo {author} {\bibfnamefont {Julio}\ \bibnamefont {{San Rom{\'a}n}}},
  \bibinfo {author} {\bibfnamefont {Antonio}\ \bibnamefont {Pic{\'o}n}},
  \bibinfo {author} {\bibfnamefont {Kevin~M.}\ \bibnamefont {Dorney}}, \bibinfo
  {author} {\bibfnamefont {Henry~C.}\ \bibnamefont {Kapteyn}}, \bibinfo
  {author} {\bibfnamefont {Margaret~M.}\ \bibnamefont {Murnane}}, \bibinfo
  {author} {\bibfnamefont {Luis}\ \bibnamefont {Plaja}}, \bibinfo {author}
  {\bibfnamefont {Maciej}\ \bibnamefont {Lewenstein}}, \ and\ \bibinfo {author}
  {\bibfnamefont {Carlos}\ \bibnamefont {Hern{\'a}ndez-Garc{\'i}a}},\
  }\bibfield  {title} {\enquote {\bibinfo {title} {{Conservation of Torus-knot
  Angular Momentum in High-order Harmonic Generation}},}\ }\href {\doibase
  10.1103/PhysRevLett.122.203201} {\bibfield  {journal} {\bibinfo  {journal}
  {Phys. Rev. Lett.}\ }\textbf {\bibinfo {volume} {122}},\ \bibinfo {pages}
  {203201} (\bibinfo {year} {2019}{\natexlab{b}})}\BibitemShut {NoStop}%
\bibitem [{\citenamefont {Rego}\ \emph {et~al.}(2019)\citenamefont {Rego},
  \citenamefont {Dorney}, \citenamefont {Brooks}, \citenamefont {Nguyen},
  \citenamefont {Liao}, \citenamefont {{San Rom{\'a}n}}, \citenamefont {Couch},
  \citenamefont {Liu}, \citenamefont {Pisanty}, \citenamefont {Lewenstein}
  \emph {et~al.}}]{rego2019generation}%
  \BibitemOpen
  \bibfield  {author} {\bibinfo {author} {\bibfnamefont {Laura}\ \bibnamefont
  {Rego}}, \bibinfo {author} {\bibfnamefont {Kevin~M}\ \bibnamefont {Dorney}},
  \bibinfo {author} {\bibfnamefont {Nathan~J}\ \bibnamefont {Brooks}}, \bibinfo
  {author} {\bibfnamefont {Quynh~L}\ \bibnamefont {Nguyen}}, \bibinfo {author}
  {\bibfnamefont {Chen-Ting}\ \bibnamefont {Liao}}, \bibinfo {author}
  {\bibfnamefont {Julio}\ \bibnamefont {{San Rom{\'a}n}}}, \bibinfo {author}
  {\bibfnamefont {David~E}\ \bibnamefont {Couch}}, \bibinfo {author}
  {\bibfnamefont {Allison}\ \bibnamefont {Liu}}, \bibinfo {author}
  {\bibfnamefont {Emilio}\ \bibnamefont {Pisanty}}, \bibinfo {author}
  {\bibfnamefont {Maciej}\ \bibnamefont {Lewenstein}},  \emph {et~al.},\
  }\bibfield  {title} {\enquote {\bibinfo {title} {{Generation of
  extreme-ultraviolet beams with time-varying orbital angular momentum}},}\
  }\href@noop {} {\bibfield  {journal} {\bibinfo  {journal} {Science}\ }\textbf
  {\bibinfo {volume} {364}},\ \bibinfo {pages} {9486} (\bibinfo {year}
  {2019})}\BibitemShut {NoStop}%
\bibitem [{\citenamefont {Groth}\ \emph {et~al.}(2014)\citenamefont {Groth},
  \citenamefont {Wimmer}, \citenamefont {Akhmerov},\ and\ \citenamefont
  {Waintal}}]{Groth_2014}%
  \BibitemOpen
  \bibfield  {author} {\bibinfo {author} {\bibfnamefont {Christoph~W}\
  \bibnamefont {Groth}}, \bibinfo {author} {\bibfnamefont {Michael}\
  \bibnamefont {Wimmer}}, \bibinfo {author} {\bibfnamefont {Anton~R}\
  \bibnamefont {Akhmerov}}, \ and\ \bibinfo {author} {\bibfnamefont {Xavier}\
  \bibnamefont {Waintal}},\ }\bibfield  {title} {\enquote {\bibinfo {title}
  {{Kwant: a software package for quantum transport}},}\ }\href {\doibase
  10.1088/1367-2630/16/6/063065} {\bibfield  {journal} {\bibinfo  {journal}
  {New Journal of Physics}\ }\textbf {\bibinfo {volume} {16}},\ \bibinfo
  {pages} {063065} (\bibinfo {year} {2014})}\BibitemShut {NoStop}%
\bibitem [{\citenamefont {Eckardt}\ and\ \citenamefont
  {Anisimovas}(2015)}]{Eckardt_2015}%
  \BibitemOpen
  \bibfield  {author} {\bibinfo {author} {\bibfnamefont {Andr{\'e}}\
  \bibnamefont {Eckardt}}\ and\ \bibinfo {author} {\bibfnamefont {Egidijus}\
  \bibnamefont {Anisimovas}},\ }\bibfield  {title} {\enquote {\bibinfo {title}
  {{High-frequency approximation for periodically driven quantum systems from a
  Floquet-space perspective}},}\ }\href {\doibase
  10.1088/1367-2630/17/9/093039} {\bibfield  {journal} {\bibinfo  {journal}
  {New Journal of Physics}\ }\textbf {\bibinfo {volume} {17}},\ \bibinfo
  {pages} {093039} (\bibinfo {year} {2015})}\BibitemShut {NoStop}%
\end{thebibliography}%
\end{document}